\documentclass{article}
\usepackage{amsfonts}
\usepackage{spconf,amsmath,graphicx,epsfig}
\usepackage{graphicx}
\usepackage{color}
\usepackage{mathtools}
\usepackage{microtype}
\usepackage{xcolor}
\usepackage{caption}
\usepackage{subcaption}
\usepackage{adjustbox}
\usepackage{lineno}
\usepackage{tabularx}
\usepackage{tikz}

\title{A Quantum Kernel Learning Approach to Acoustic Modeling\\ for Spoken Command Recognition}

\name{\begin{tabular}{@{}c@{}}
Chao-Han Huck Yang$^{1,2*}$\thanks{$^*$Work on acoustic feature encoding was done at Google as an intern.}, Bo Li$^{2}$, Yu Zhang$^{2}$, Nanxin Chen$^{2}$, Tara N. Sainath$^{2}$\\ Sabato Marco Siniscalchi$^{1,3,4}$, Chin-Hui Lee$^{1}$
\end{tabular}}
\address{$^1$ Georgia Institute of Technology, USA \qquad $^2$Google, USA \\$^3$Kore University of Enna, Italy \qquad $^4$Department of Electronic Systems, NTNU, Trondheim, Norway }

\usepackage{hyperref}

\begin{document}

\maketitle
\begin{abstract}
We propose a quantum kernel learning (QKL) framework to address the inherent data sparsity issues often encountered in training large-scare acoustic models in low-resource scenarios. We project acoustic features based on classical-to-quantum feature encoding. Different from existing quantum convolution techniques, we utilize QKL with features in the quantum space to design kernel-based classifiers. Experimental results on challenging spoken command recognition tasks for a few low-resource languages, such as Arabic, Georgian, Chuvash, and Lithuanian, show that the proposed QKL-based hybrid approach attains good improvements over existing classical and quantum solutions.

\end{abstract}
\section{Introduction}\label{sec:introduction}
Recently, deep neural network~\cite{hinton2012deep, sainath2015convolutional} (DNN) models have demonstrated a competitive performance on many speech processing tasks. Nonetheless, training a large parameterized DNN using as few as 1,000 utterances usually leads to a poor speech recognition accuracy. Considering that there exist over $8,000$ spoken languages~\cite{lewis2009ethnologue} in the world, it is clear that some of those spoken languages may not provide enough training materials to properly deploy DNN-based spoken command recognition (SCR) systems. Meanwhile, we have witnessed a rapid growth of quantum devices, and quantum machine learning (QML) systems can now ``be deployed in practices'' thanks to software simulator (e.g., Tensorflow-quantum~\cite{broughton2020tensorflow}). QML-based algorithms can also be combined with modern DNN solutions to accomplish feature extraction or reduce computation complexity. In the present quantum computing era~\cite{boixo2018characterizing} in which $4$ to $200$ qubits are accessible, an integration of quantum based components (e.g., feature extractor) into a classical ML module, such as quantum neural networks~\cite{abbas2021power, killoran2019continuous, qi2022classical} (QNNs) and quantum kernel learning~\cite{havlivcek2019supervised} (QKL), is a realistic solution that can be deployed for real-world applications. QNNs use parameterized quantum circuits (e.g., tensor network) to optimize latent features~\cite{havlivcek2019supervised} in the Hilbert space~\cite{cybenko1989approximation} for minimizing a loss function. QNNs can be trained under gradient optimization with variational circuits~\cite{mitarai2018quantum} in an end-to-end fashion to approximate any non-linear~\cite{schuld2021supervised} function.

\begin{figure}[ht!]
\begin{center}
\vspace{-2mm}
  \includegraphics[width=0.95\linewidth]{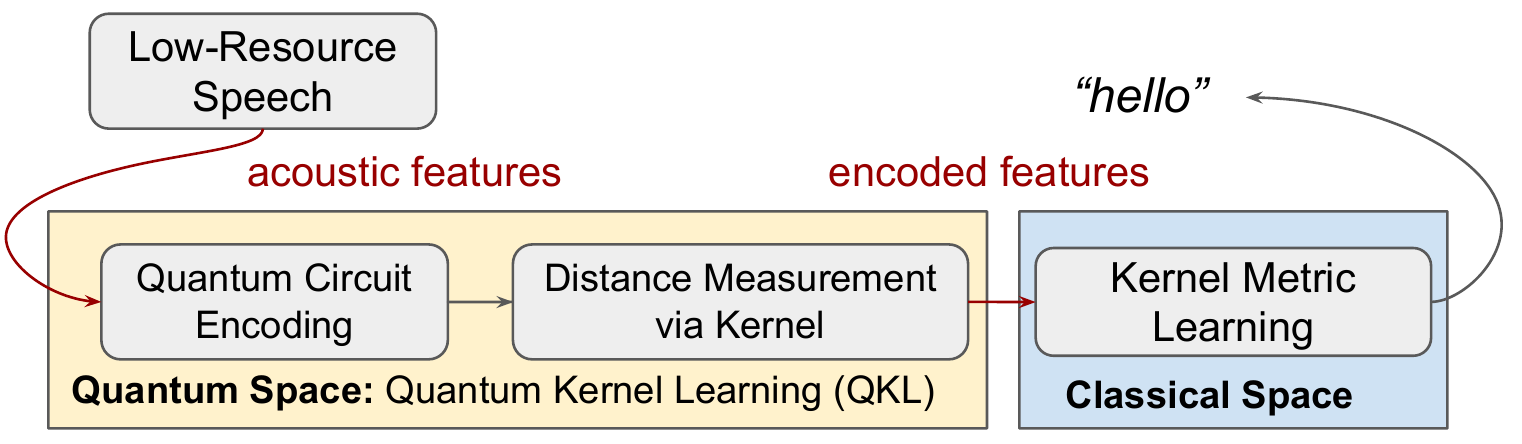}
\end{center}
\vspace{-0.2cm}
  \caption{Computing diagram of quantum kernel learning, where quantum kernel projection occurs on near-term quantum processing units (QPU) or simulators (e.g., CPU or TPU).
  }
\label{fig:0}
\end{figure}
\vspace{-3mm}

However, such parameterized variational quantum algorithms are often associated with problems of ``barren plateaus''~\cite{grant2019initialization, mcclean2018barren}, where training loss takes an extremely long time to update through gradients. QKL~\cite{huang2021power, schuld2021supervised} instead provides an alternative training mechanism to use quantum states for quantum encoding and projection~\cite{nielsen2002quantum}, and then forms a kernel to estimate a hyperplane to separate training data with its kernel alignment. We will discuss these quantum operations in detail later. Recent theoretical studies~\cite{schuld2021supervised, schuld2019quantum} have also proven that QKL requires less trainable parameters to archive the same performance of QNNs. These findings could potentially tackle the current challenges in low-resource SCR.  In this paper, we propose a quantum kernel projection \footnote{A tutorial and guidelines for QKL can be found in~\url{https://www.tensorflow.org/quantum/tutorials/quantum_data}.} based QML scheme that uniformly projects acoustic features in a high-dimensional space during its classical-to-quantum feature encoding phase to establish hybrid quantum-classical classification models. Figure~\ref{fig:0} provides an example showing how to encode input acoustic features into quantum states and to capture latent representations through quantum projections in the left block for kernel metric learning~\cite{weinberger2007metric, li2007approximate, pargellis2001metrics} in the right block of Figure~\ref{fig:0}. Note that quantum computing processors can be deployed in software-simulated toolkit~\cite{broughton2020tensorflow,bergholm2018pennylane} and noisy intermediate-scale quantum~\cite{preskill2018quantum} (NISQ) hardware that are currently accessible with $1$ to $100$ qubits.

\section{Related Work}\label{sec:related_work}
\subsection{Low-Resources Spoken Command Recognition}
Several competitive techniques have been developed for recognizing spoken commands in various voice assistant applications (e.g., intelligent home control). On the one hand, DNN-based acoustic models often achieve over $90\%$ accuracies for resource-sufficient languages (e.g., English~\cite{Warden2018}). On the other hand, statistical learning (e.g., kernel SVM~\cite{cortes1995support}) has been proven to be a viable solution for low-resource languages~\cite{karunanayake2019transfer}, such as Sinhala and Tamil. Moreover, prototypical network based metric learning~\cite{huh2021metric, chung2020defence} shows competitive performances in few-shot speaker identification tasks~\cite{chung2020defence,qian2021speech, tian2020improving}, where the model~\cite{huh2021metric} learns per-class weights for target keywords, as in classification objectives and maximizes the distance between non-targeted samples.
In this work, we will show that such an efficient non-parametric learning approach can be combined with quantum computation for low-resource spoken command recognition.

\subsection{Applications of Quantum Machine Learning}
Under the current computing constraints of limited qubits, quantum machine learning (QML) has been mainly deployed with quantum circuits~\cite{mitarai2018quantum} for some classical image or audio processing problems. The existing QML models are often hybrid quantum-classical models that use quantum circuits as a parameterized encoder or decoder. Those quantum circuits project the input features into a high-dimension space and rotate the quantum state representations. Some preliminary studies on quantum circuit based convolution~\cite{yang2021decentralizing} showed that hybrid models could work effectively for English spoken command recognition. However, the training time of a parameterized circuit usually relies on the required variational gradient approximation algorithms~\cite{caro2022generalization}. Instead, a quantum kernel technique provides another perspective of using the characterises of quantum space on non-parametric learning, which could be beneficial to the few-shot and low-resource training scenarios.

\section{Quantum Projection for Kernel Learning}
\subsection{Quantum Feature Encoding and Projection}

Unlike variational circuit learning~\cite{mitarai2018quantum} that requires the number of model parameters in the order of the training set size, quantum kernel learning (QKL)~\cite{huang2021power} can be adopted for limited training scenarios. QKL builds a quantum mapping, $\phi$, transforming a classical input vector $\textbf{x}$ into a quantum state $\vert \phi(\textbf{x})\rangle$ in a quantum Hilbert space.

Let $Q$ stands for the total number of qubits in the utilized quantum circuit and $\vert \delta_{q} \rangle$, $q=1,...,Q$, stands for the quantum state of the $q^{th}$ qubit with two possible values, $\delta_q =0$ or $\delta_q = 1$ in Dirac notations~\cite{nielsen2002quantum}. $\vert 0 \rangle^{\otimes Q}$ refers to the \textbf{tensor product} of $\vert 0 \rangle$ of $Q$ times, which is equivalent to $\vert 0 \rangle \otimes \vert 0 \rangle \otimes \cdot\cdot\cdot \otimes \vert 0 \rangle$, as the initial quantum state. More specifically, we consider a quantum encoding by using a parametric quantum circuit, with a unitary operator $U(\textbf{x})$ of $Q$ qubits, such that the input vector $\textbf{x}$ is encoded into an embedded quantum state for the entire circuit, $\vert \phi(\textbf{x}) \rangle$. In doing so, $\vert \phi_q(\textbf{x}) \rangle$ represents the quantum state of $q^{th}$ qubit after encoding.
Next, we take a quantum measurement of $\vert \phi(\textbf{x})\rangle$ based on the Pauli-Z axis for many times \cite{nielsen2002quantum}, and then averages the measured values to produce a vector output, $\textbf{y}$, with the same dimension of the original input vector $\textbf{x}$. Now we can project the associated embedded quantum states $\vert \phi(\textbf{x}) \rangle$ back to the original classical space, such that for each pair ($\textbf{y}_{i}, \textbf{y}_{j}$), a kernel can be redefined as:
\begin{align}
K_{C}(\textbf{y}_{i}, \textbf{y}_{j}) \rightarrow~&K_{Q}(\textbf{x}_{i}, \textbf{x}_{j}) = \left|\langle \phi(\textbf{x}_{i}) \vert \phi(\textbf{x}_{j}) \rangle \right|,
\label{eq:6}
\end{align}
    
where $K_C(\textbf{y}_{i}, \textbf{y}_{j})$ represents the kernel in a classical sense while $K_{Q}(\textbf{x}_{i}, \textbf{x}_{j})$ denotes quantum kernels. Finally, we calculate the inner-product of the two quantum feature maps, $\phi(\textbf{x}_{i})$ and $\phi(\textbf{x}_{j})$, in Eq.~(\ref{eq:6}) as a distance measure for kernel learning, which are computed by first-order reduced density matrix $\rho(\textbf{x})$ of classical input vector $\textbf{x}$ with circuit encoding. The actual measurement~\cite{hubregtsen2022training} between $\phi(\textbf{x}_{i})$ and $\phi(\textbf{x}_{j})$ is formed as a Gaussian kernel projection, $K_{QG} = \exp \left(-\gamma \sum_k\left(\operatorname{Tr}\left[\rho\left(\textbf{x}_{i}\right)\right]-\operatorname{Tr}\left[\rho\left(\textbf{x}_{j}\right)\right]\right)^2\right)$, where $\operatorname{Tr[\cdot]}$ is the trace operator and $\gamma$ is set to be $1$ in our study.

\subsection{Metric Learning with Quantum Measurements}
We use Gaussian QKL combined with one competitive metric learning solution of prototypical networks~\cite{snell2017prototypical, huh2021metric}. Our training objective is jointly built upon generalised end-to-end loss~\cite{wan2018generalized} and angular variant prototypical loss used in~\cite{huh2021metric} for training with low-resource speech data. For model training, we first collect latent feature embeddings projected from output measurement of quantum kernel and use prototypical loss to maximize class-wise distance in the latent space. We train a prediction backbone with kernel SVM referred to the setup in~\cite{huh2021metric}. We then re-use the quantum projection with kernel learning in Eq.~(\ref{eq:6}) from the NISQ device for kernel SVM learning as a hybrid model. Similar to the findings in~\cite{huh2021metric}, kernel SVM performs better than DNN serving a backbone prediction model evaluated in our low-resource command data.

\section{Experiments and Result Analysis}\label{sec:experiment}

\subsection{Experimental Setup and Baselines}
\label{sec:4:1:data}
\begin{figure}[ht!]
\begin{center}
\vspace{-2mm}
  \includegraphics[width=0.90\linewidth]{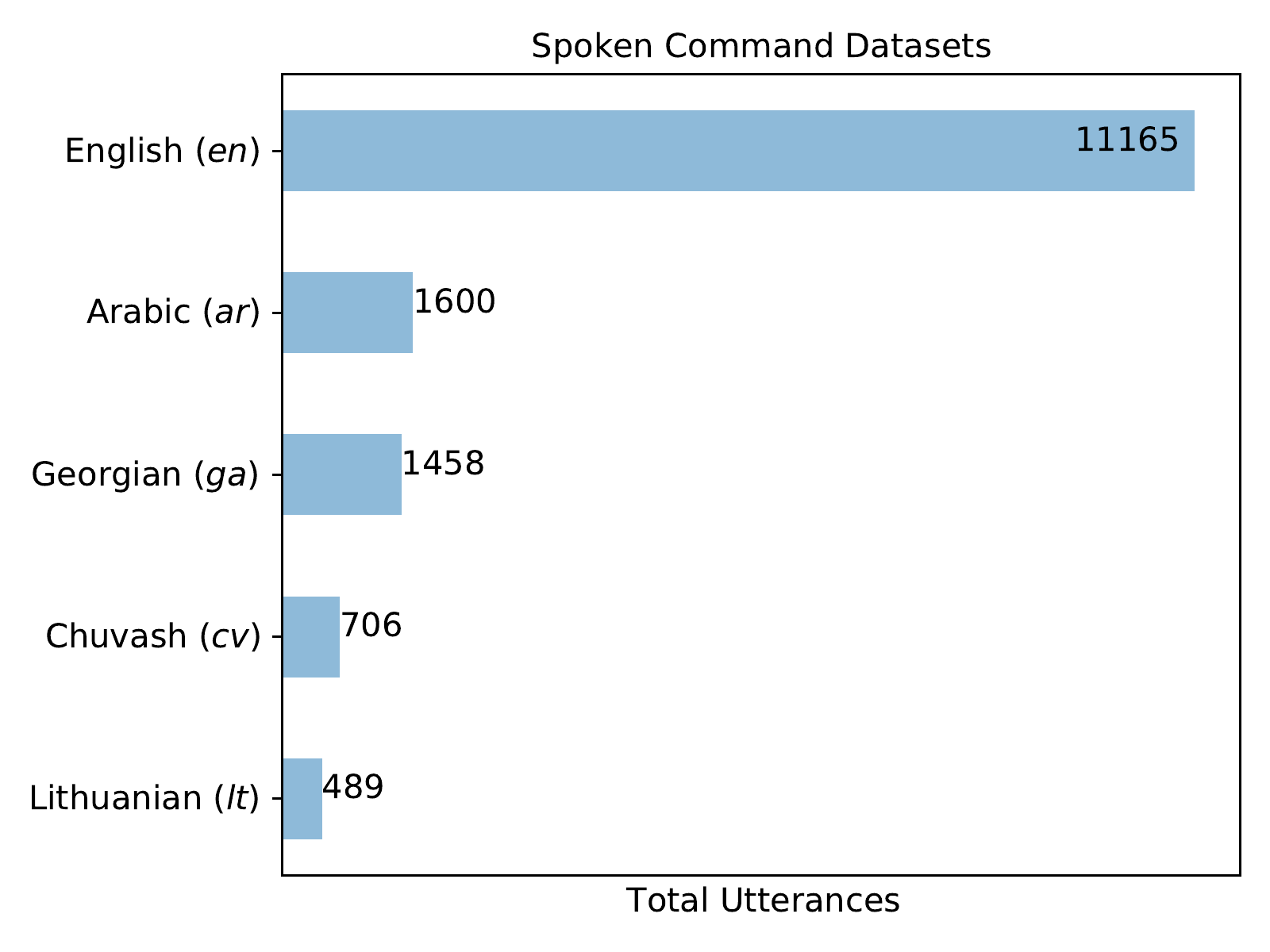}
\end{center}
  \caption{Training set sizes for 5 languages used here: Arabic, Georgian, Chuvash, and Lithuanian are considered low-resource when compared to English (with $\sim11$k utterances).
  }
\label{fig:data}
\end{figure}

We first test the English ($\mathbf{en}$) Google Speech Command~\cite{Warden2018} corpus, containing a total of up to $11,165$ training and $6,500$ testing utterances in $1$-second length. Next, for the low-resource spoken command recognition task we have four languages: Georgian ($\mathbf{ga}$) with $1458$ utterances and Chuvash ($\mathbf{cv}$) with $706$ utterances are collected from the Mozilla Common Voice data set~\cite{ardila2020common} with the same 10 frequent commands translating from the above set of English words. Meanwhile, Arabic ($\mathbf{ar}$)~\cite{Benamer2020} and Lithuanian~\cite{Kolesau2020} ($\mathbf{lt}$) Speech Commands are referred to the setup in the existing spoken command classification studies. Arabic includes $1600$ utterances with $6$ smart home control words and 10 spoken digits ($0$ through $9$). For Georgian, Chuvash, and Arabic data, we use the same setup as in~\cite{Benamer2020} to split these sets into $80$\% for training and $20\%$ for testing with $10$-fold cross-validation. Lithuanian duses $326$ , $75$ and $88$ utterances for training, validation and testing, respectively. In Figure~\ref{fig:data} we display the maximum training set sizes for each language used. Clearly there is a noticeable difference between English and the other four low-resource languages.

\subsubsection{Data Pre-processing}
We follow the data-format used in Google Speech Command~\cite{Warden2018} and refine all input utterances of low-resource spoken commands to under $1$ second. We use the original white noise setup in~\cite{Warden2018} to simulate a noisy background. Mel-scaled spectra in $60$ bands and $1024$-point discrete Fourier transform are extracted from the input speech signals have been extracted its with Librosa~\cite{mcfee2015librosa} and Keras audio pre-processing layer~\cite{choi2017kapre}. For NISQ simulation, We run TensorFlow quantum on TPU for our experiments with 22.7\% reduced decoding time compared to CPU-based simulation.

\subsubsection{Baseline Algorithms}
We select three competitive spoken command classification models of ``pure DNN'', ``QCNN-DNN'', and ``kernel metric learning'' as baselines. Following the setup in~\cite{yang2021decentralizing}, we first select an attention recurrent neural network~\cite{de2018neural}, as our ``pure DNN'' baseline. We next select its extended hybrid quantum-classical version (denoted as ``QCNN-DNN``) proposed in~\cite{yang2021decentralizing}, where a quantum convolution layer has been used to replace some of the neural convolution layers of a pure DNN model with a boosted performance. We use the best setup of $2\times2$ convolution filter with $4$ qubits in~\cite{yang2021decentralizing} (increasing the qubits numbers up to $9$ does not improve performances from our previous study~\cite{yang2021decentralizing}). Finally, for the "kernel metric learning" baseline, we use the best setup for speech command classification proposed in~\cite{huh2021metric} with a backbone of multi-class radial basis function (RBF) based ``kernel SVM''~\cite{vert2004primer} using a typical loss ~\cite{snell2017prototypical} to optimize its latent embedding as a distance metric.

\subsection{Classification Result and Analysis} \label{sec:result}
\subsubsection{A Preliminary Study on Using Kernels}
First, we did a quick study to compare QKL with other kernel-based metric learning with an inner-product cosine distance, and display the results in Table~\ref{tab:kernel}, where we randomly select $1,000$ English speech data for evaluation on its validation accuracy and clustering distance between each class.  As a result, Gaussian-based quantum projection shows the best accuracy and smallest distance compared to quantum linear kernel and metric learning~\cite{huh2021metric} baselines.
\vspace{-0.5\baselineskip}
\begin{table}[ht!]
\centering
\caption{Comparing different quantum kernel learning setups}
\label{tab:kernel}
\begin{tabular}{|l|c|c|}
\hline
Method & Acc. ($\uparrow$) & Cluster Dist. ($\downarrow$) \\ \hline
Kernel metric learning~\cite{huh2021metric} & 65.3\% & 2.12 \\ \hline \hline
Linear-QKL & 66.4\%&  2.02 \\ \hline
Gaussian-QKL & \textbf{72.6\%}& \textbf{1.83} \\ \hline
\end{tabular}
\end{table}

\subsubsection{English Spoken Command Recognition} 
To investigate the impact of insufficient training data, we train different classification models introduced in Sec.~\ref{sec:4:1:data} from scratch and report their average accuracies under $10$-fold cross-validation. As shown in Table~\ref{tab:eng}, DNN-based baselines outperform the two other baselines and QKL, from $1.7$\% to $4.8$\% in large-scale training with $11$k training utterances. However, QKL demonstrates the best prediction performance after the training data scale down to 1k and continuously performs as the best compared to all three other baselines even with a limited training set of $500$ utterances. As a preliminary finding, DNN-based models show a serve $57.1$\% accuracy drop from $95.2$\% with full training (with $11$k utterances) to 38.1\% in low-resource training (with only $500$ samples).
\begin{table}[ht!]
\centering
\caption{Comparing average classification accuracies (in \%) of English Speech Command under different training set sizes. }
\label{tab:eng}
\begin{adjustbox}{width=0.48\textwidth}
\begin{tabular}{|l|l|l|l|l|}
\hline
Number of training Utterances & 500 & 1k & 5k & 11k \\ \hline \hline
Pure DNN~\cite{de2018neural} & 38.1 & 57.2 & 83.6 & \textbf{95.2} \\ \hline
Kernel metric learning~\cite{huh2021metric} & 44.4 & 61.8 & 77.4 & 90.4 \\ \hline \hline
QCNN-DNN~\cite{yang2021decentralizing} & 41.2 & 62.9 & 80.5 & 93.2 \\ \hline
Gaussian-QKL (proposed) & \textbf{47.7} & \textbf{72.6} & \textbf{84.1} & 93.5 \\ \hline
\end{tabular}
\end{adjustbox}
\end{table}

\subsubsection{Low-resource Spoken Command Recognition} 
In Table~\ref{tab:mul}, we display classification accuracies for the four low-resource languages, $\{\mathbf{ga}$, $\mathbf{cv}$, $\mathbf{lt}$, $\mathbf{ar}\}$ discussed in Sec.~\ref{sec:4:1:data}. Based on the evaluation results, $\mathbf{cv}$ is the most challenging language with a low accuracy ($<45\%$) on all models. DNN (the third row of Table~\ref{tab:mul}) shows poor accuracies of $17.6$\% and $46.3$\% on the $\mathbf{cv}$ and $\mathbf{lt}$ when training data are fewer than $1$k samples. As a hybrid quantum-classical model, QCNN-DNN demonstrates slightly better performances than its DNN counterparts on the $\mathbf{ga}$, $\mathbf{cv}$, and $\mathbf{ar}$ test sets. The proposed QKL-based models (the sixth row of Table~\ref{tab:mul}) perform the best when compared to the other baselines in all four evaluated low-resource languages.

\begin{table}[ht!]
\centering
\caption{A comparison of average classification accuracies (in \%) for four low-resource languages: Georgian ($\mathbf{ga}$), Chuvash ($\mathbf{cv}$), Lithuanian ($\mathbf{lt}$), and Arabic ($\mathbf{ar}$).}
\label{tab:mul}
\begin{tabular}{|l|c|c|c|c|}
\hline
Language& $\mathbf{ga}$ & $\mathbf{cv}$ & $\mathbf{lt}$ & $\mathbf{ar}$ \\ \hline
\# Total classes & $10$ & $10$ & $15$ & $16$ \\ \hline \hline
Pure DNN~\cite{de2018neural} & 56.4 & 17.6 & 46.3 & 66.4 \\ \hline
Kernel metric learning~\cite{huh2021metric} & 57.3 & 30.7 & 44.2 & 63.6 \\ \hline
QCNN-DNN~\cite{yang2021decentralizing} & 58.5 & 28.9 & 45.9 & 67.2 \\ \hline
Gaussian-QKL (proposed) & \textbf{75.1} & \textbf{41.5} & \textbf{57.9} & \textbf{70.4} \\ \hline
\end{tabular}
\end{table}

\subsection{Discussions} \label{sec:discussion}
\subsubsection{Performances versus Training Epochs} 
\begin{figure}[ht!]
\begin{center}
  \includegraphics[width=0.80\linewidth]{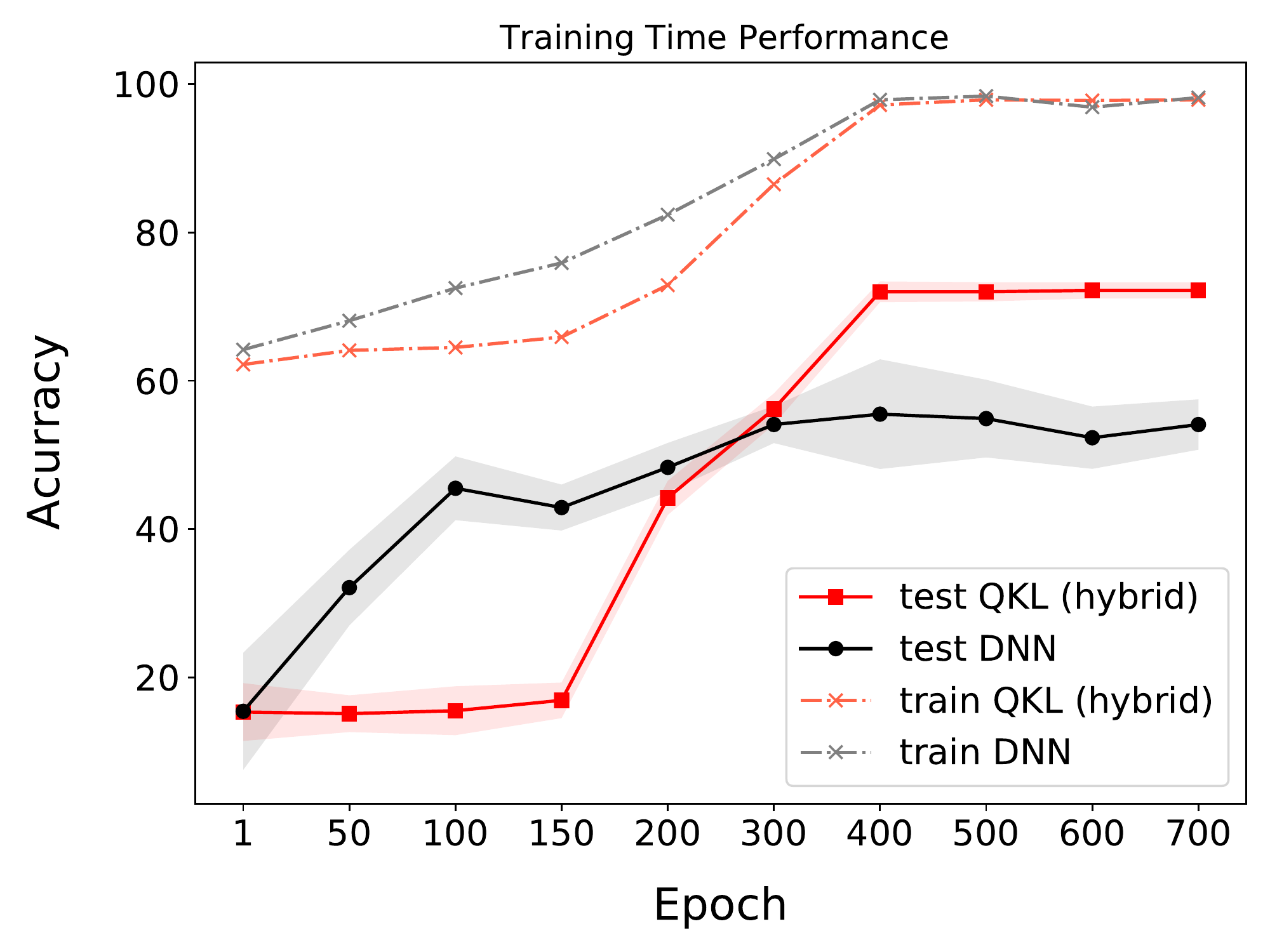}
\end{center}
\vspace{-0.2cm}
  \caption{Average training and test set accuracies at differnt training epochs evaluating on four low-resource languages.
  }
\label{fig:tr}
\end{figure}

As the first time to report spoken command classification with quantum kernel-based learning, we further report its training time behavior as shown in Figure~\ref{fig:tr} when compared to existing end-to-end training with the DNN-based models.

Interestingly, QKL models (\textcolor{red}{in red curves}) take more training epochs than DNN models (in black curves) to attain an accuracy of over $40$\%. Convergence of DNN-based models seems to be faster in both training and testing sets. This phenomenon could be related to the discussion of ``barren plateaus'' in previous QML studies~\cite{grant2019initialization, mcclean2018barren}. In summary, both DNN and QKL models seem to be over-fitted at the training set as shown in the two upper dash-line curves. However, the test set performance of QKL shows a better than $15$\% accuracy when compared to the test set performance of DNN when comparing the two lower solid-line curves.

\subsubsection{Low-resource SCR with Different Training Set Sizes} 
\begin{figure}[ht!]
\begin{center}
  \includegraphics[width=0.83\linewidth]{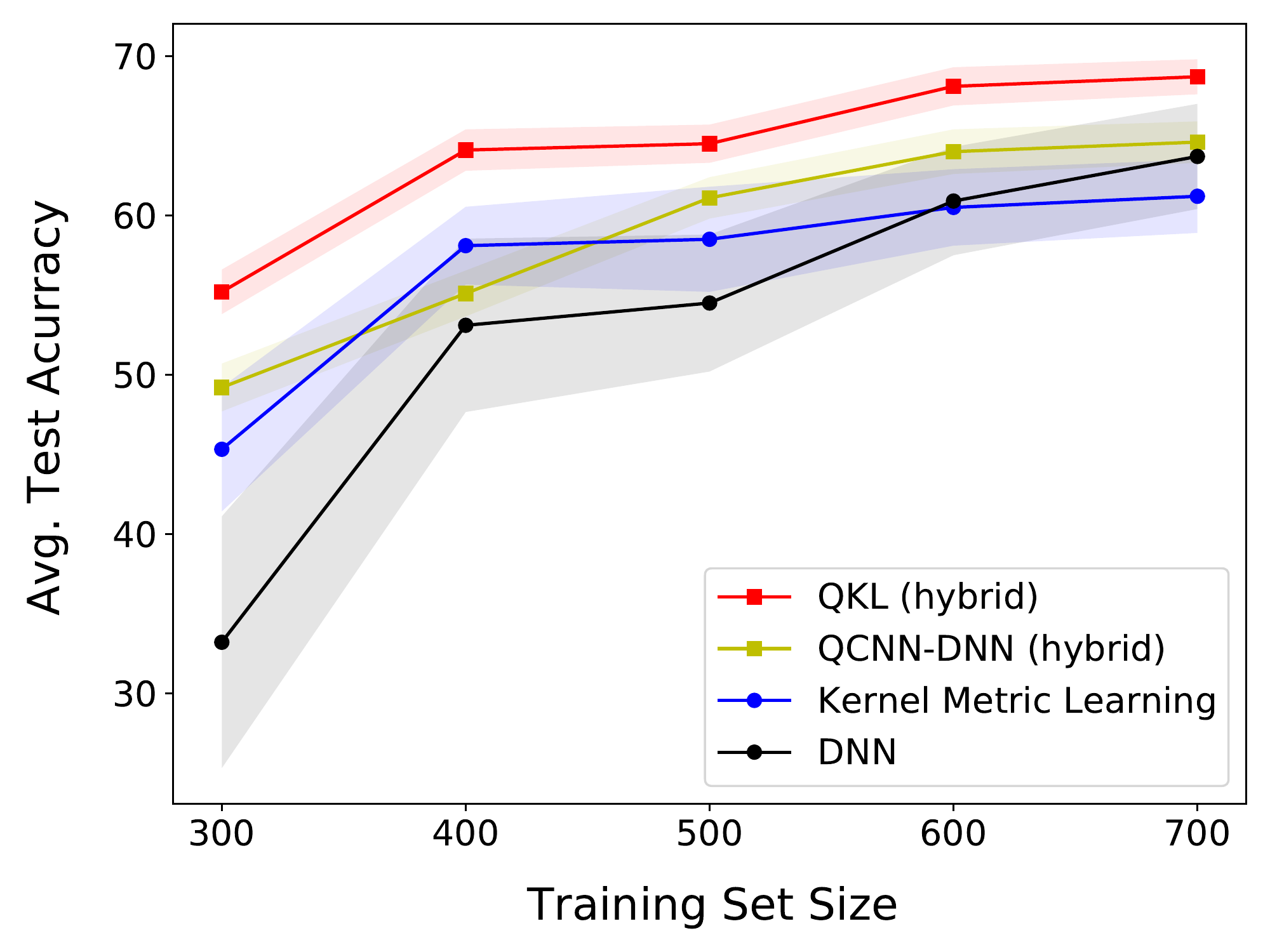}
\end{center}
\vspace{-0.2cm}
  \caption{A comparison of average test accuracy of kernel learning for low-resource SCR in different training set sizes.
  }
\label{fig:1}
\end{figure}

Finally, we conduct an additional evaluation of low-resource spoken commands under different training set sizes ranging from $300$ to $700$, for $\mathbf{ga}$, $\mathbf{cv}$, and $\mathbf{ar}$. We exclude the $\mathbf{lt}$ data since its maximum training set size is just about $300$. The data in the three languages have been scaled down to observe the sensitivity of each trained model. We report average validation accuracies (in the $y$-axis of Figure~\ref{fig:1}) weighted by the total number of samples of each of the three evaluated languages reported in Figure~\ref{fig:data}. Then , based on Figure~\ref{fig:tr}, we notice QKL continuously outperforms the other setups with better accuracies. QKL also shows a more stable performance in terms of convergence rate when compared to the other algorithms.

\vspace{-3mm}

\section{Summary}
We propose kernel-based learning for classifying low-resource spoken commands. Our experimental results suggest that: (i) DNN-based acoustic models are sensitive to training set sizes, and (ii) quantum kernel learning is effective in classifying low-resource spoken commands. QKL-based learning also gives more stable and better validation accuracies when compared to existing kernel learning and hybrid quantum-DNN models.

\clearpage
\footnotesize
\bibliographystyle{IEEEtran}
\bibliography{ref}

\end{document}